\documentclass[twocolumn,prb,tightenlines,superscriptaddress,amsmath,amssymb,amsfonts,noeprint]{revtex4-2}
\usepackage{graphicx}
\usepackage{dcolumn}
\usepackage{bm}
\usepackage{float}
\usepackage{hyperref}
\usepackage[section]{placeins}
\usepackage{epstopdf}
\usepackage[table, dvipsnames]{xcolor}
\usepackage[T1]{fontenc}
\usepackage{times}
\usepackage{charter}
\usepackage[expert]{mathdesign}

\newcommand{\ie}{\textit{i}.\textit{e}.}

\begin{document}
\title{Instantaneous Emission Rate of Electron Transport through a quantum point contact}
\author{Y. Yin}
\thanks{Author to  whom correspondence should be addressed}
\email{yin80@scu.edu.cn.}
\affiliation{Department of Physics,
  Sichuan University, Chengdu, Sichuan, 610065, China}
\date{\today}
\begin{abstract}
  We present a theory to describe the instantaneous emission rate of electron
  transport in quantum-coherent conductors. Due to the Pauli exclusion
  principle, electron emission events are usually correlated. This makes the
  emission rate is not a constant, but depends on the history of the emission
  process. To incorporate the history dependence, in this paper we characterize
  the emission rate via the conditional intensity function, which has been
  introduced in the theory of random point process. The conditional intensity
  function can be treated as the instantaneous emission rate observed by an
  ideal single-electron detector. We demonstrate the method by studying the
  instantaneous emission rate of a single-channel quantum point contact driven
  by a constant voltage. As the quantum point contact is opened up, we show that
  the emission process evolves from a simple Poisson process close to pinch-off
  to a non-renewal process at full transmission. These results show that the
  conditional intensity function can provide an intuitive and unified
  description of the emission process in quantum-coherent conductors.
\end{abstract}
\pacs{73.23.-b, 72.10.-d, 73.21.La, 85.35.Gv}
\maketitle

\section{Introduction}
\label{sec1}

The electron emission in quantum conductors is an inherently stochastic process,
which has been extensively studied for several decades \cite{Levitov1996,
  Blanter1999}. In a typical setup, electrons are emitted from the reservoir
into the conductor through a quantum point contact (QPC), which are driven by a
constant bias voltage $V$. As the QPC is close to pinch-off, the emission events
are rare and nearly uncorrelated. The absence of the correlation can be seen
from the dc shot noise, which follows the Schottky formula
$S_{Poisson} = 2 e \bar{I}$, with $\bar{I}$ being the dc current and $e$ being
the electron charge. It indicates that the electron emission can be described by
a simple Poisson process with a constant emission rate
$\bar{\lambda} = \bar{I}/e$. This picture can be further justified from the
corresponding waiting time distribution (WTD), which can be well-approximated by
an exponential distribution
$\mathcal{W}(\tau) = \bar{\lambda} e^{-\bar{\lambda} \tau}$ \cite{Brandes2008,
  Albert2012, Haack2014, Hofer2015}.

What is the emission rate when the QPC is opened up? This is a nontrivial
question, as the emission of electrons can be correlated in this case. The
correlation can reduce the dc shot noise below the Poisson value $S_{Poisson}$,
indicating that the emission process is more regular than the Poisson process. A
more detailed information on the correlation can be obtained via the WTD. It has
been shown that the WTD can exhibit a cross-over from the exponential
distribution close to pinch-off to the Wigner–Dyson distribution at full
transmission \cite{Albert2012, Haack2014}. This shows that the emission of an
electron can be strongly hindered by the previous emitted one. In fact, the
correlation is not restricted between the two successively emitted electrons
\cite{Ptaszynski2017a, Rudge_2019, Kleinherbers2021}. Due to the Pauli exclusion
principle, the correlation is present whenever the wave functions of two emitted
electrons are overlapped in time domain. This can lead to correlations between
waiting times, which has been revealed from the joint WTD analysis
\cite{Dasenbrook2015}. Due to the correlation effects, the emission rate cannot
be a constant, but should depend on the whole history of the emission
process. As far as the author knows, the details of the emission rate has not
been fully addressed yet.

\begin{figure}
  \centering
  \includegraphics[width=7.5cm]{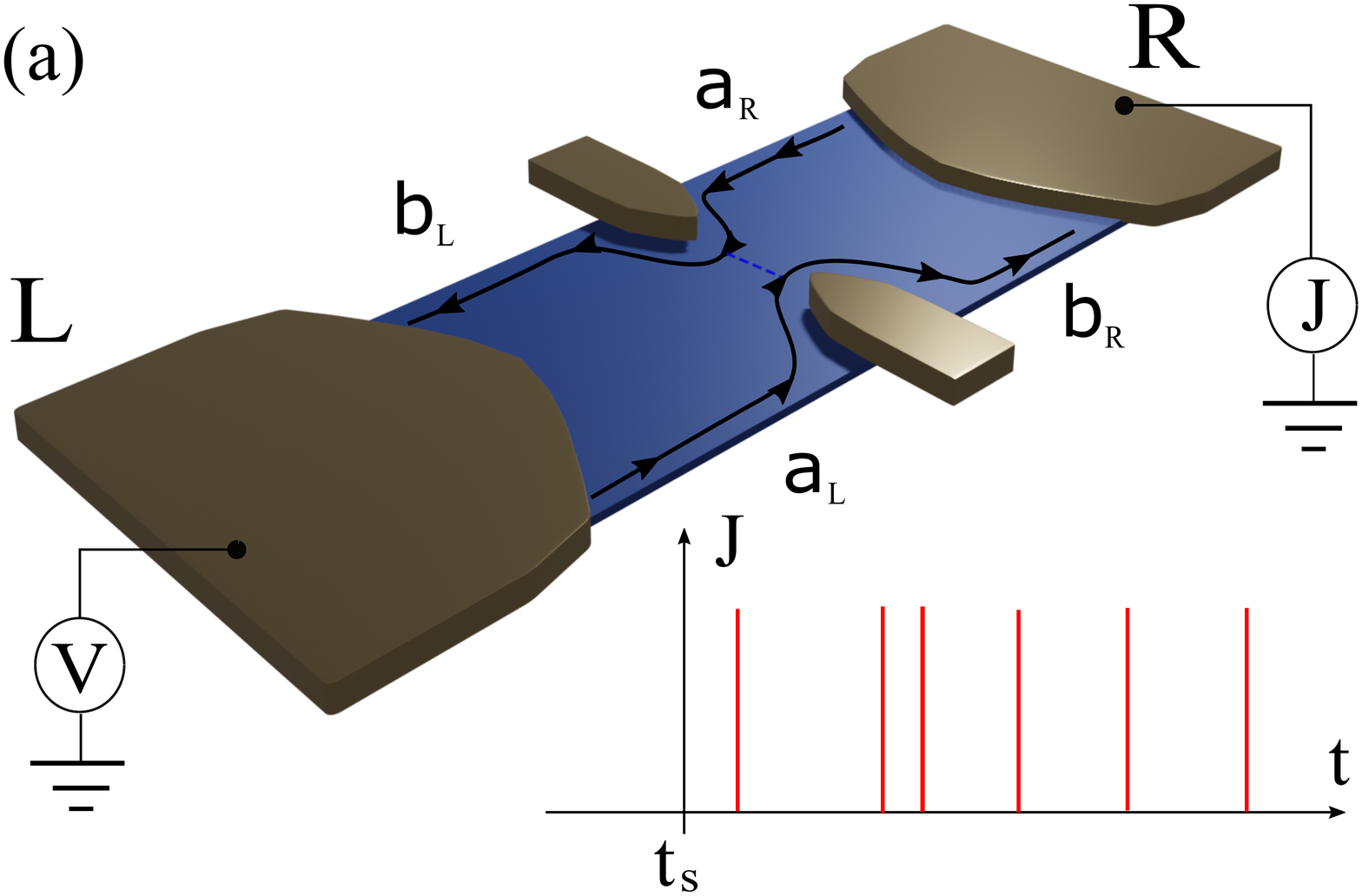}
  \includegraphics[width=8.0cm]{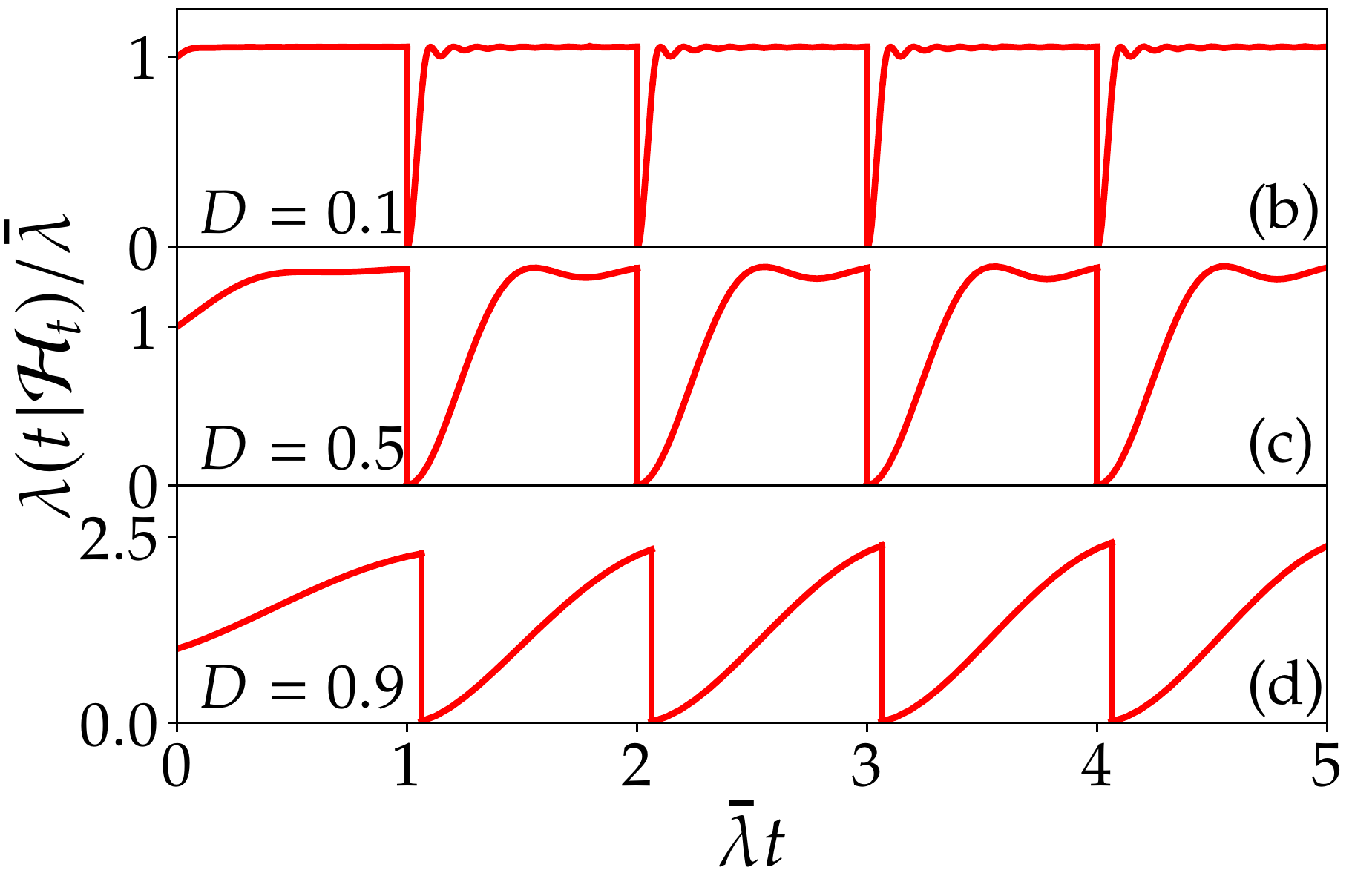}
  \caption{ (a) A quantum point contact connected to two reservoirs $L$ and
    $R$. Electrons are driven from $L$ to $R$ via a constant bias voltage
    $V$. Individual electron emission events are detected via a single-electron
    detector. The instantaneous emission rate obtained from the single-electron
    detector can be represented by the conditional intensity function
    $\lambda(t|\mathcal{H}_t)$. (b-d) Conditional intensity function
    $\lambda(t|\mathcal{H}_t)$ as a function of time $t$ for different
    transparency $D$, with $\bar{\lambda} = DeV/h$ being the average emission
    rate. Both $\lambda(t|\mathcal{H}_t)$ and $t$ are rescaled according to the
    average emission rate $\bar{\lambda}$.}
  \label{fig1}
\end{figure}

To answer this question, in this paper we characterize the emission rate via the
{\em conditional intensity function} $\lambda(t|\mathcal{H}_t)$. It is
essentially a rate function, which represents the instantaneous emission rate at
the time $t$ given the history $\mathcal{H}_t$ of the emission process
\cite{D.J.Daley2003}. The history $\mathcal{H}_t$ can be represented by an
ordered time sequence
$\mathcal{H}_t = \{ t_1, t_2, \dots : t_s < t_1 < t_2 < \dots < t \}$, where
each $t_i$ ($i=1, 2, \dots$) represents the time instant of an electron emission
event that occurs after a given time $t_s$. The time $t_s$ can be understood as
the starting time of a single-electron detector, which can record current pulses
due to individual electron emission events, as illustrated in
Fig.~\ref{fig1}(a). Due to the probabilistic nature of the electron emission,
$t_i$ are essentially random parameters. Their waiting times $\tau=t_i-t_{i-1}$
follow the WTD $\mathcal{W}(\tau)$, whose mean value is equal to the average
waiting time $1/\bar{\lambda}$ with $\bar{\lambda} = DeV/h$ being the average
emission rate.

We demonstrate the method by studying the instantaneous emission rate of a
single-channel QPC driven by a constant voltage $V$, as illustrated in
Fig.~\ref{fig1}(a). In this setup, the emission process is stationary. Hence the
instantaneous emission rate is independent of $t_s$, which can be chosen as
$t_s=0$. The typical behaviors of $\lambda(t|\mathcal{H}_t)$ for different QPC
transparency $D$ are demonstrated in Fig.~\ref{fig1}(b-d). In order to make the
discussion concrete, we assume the history $\mathcal{H}_t$ corresponds to a
sequence of equally spaced time instants, \ie,
$\mathcal{H}_t = \{ t_1 = 1/\bar{\lambda}, t_2 = 2/\bar{\lambda}, \dots : 0 <
t_1 < t_2 < \dots < t \}$. We find that $\lambda(t|\mathcal{H}_t)$ exhibits
discontinuous jumps in all cases: They drop abruptly to zero whenever $t=t_i$,
indicating the suppression of the emission rate due to the Pauli exclusion
principle. After the suppression, the emission rate start to increase, which
exhibits different behaviors for different transparency $D$.

In the case of low transparency [Fig~\ref{fig1}(b)], the emission rate increases
rapidly and saturates to the average emission rate $\bar{\lambda}$. In this
case, the correlations between emission events manifest themselves as sharp
dips, which can only play a role on short time scales. This makes the emission
process can be treated as a simple Poisson process on long time scales, which
can be fully characterized by the average emission rate $\bar{\lambda}$. For the
QPC with a modest transparency [Fig~\ref{fig1}(c)], the increasing of the
emission rate is relatively slow. It can reach the saturation value, which is
larger then the average emission rate $\bar{\lambda}$. In this case, the dips in
the emission rate evolve into wide valleys, indicating that the correlations are
non-negligible even on long time scales. Moreover, we find that the
$\lambda(t|\mathcal{H}_t)$ does not depend on the whole history, but is most
sensitive to the time instant of the emission of the previous electron. This
makes the emission process can be treated approximately as a renewal process,
where the correlations are restricted between the two successively emitted
electrons. For the QPC with high transparency [Fig~\ref{fig1}(d)], the emission
rate increases almost linearly as a function of $t$ before the saturation
occurs. The saturation value is much larger than the average emission rate
$\bar{\lambda}$ and usually cannot be reached in typical cases. As a
consequence, the emission rate exhibits a saw-tooth behavior in the time domain,
indicating the presence of strong correlations. In this case, the renewal
approximation breaks down and the emission process can only be described within
the non-renewal theory \cite{Dasenbrook2015, Ptaszynski2017, Kosov2017,
  Rudge_2019}. These results demonstrate that the conditional intensity function
can provide an intuitive and unified description of the electron emission
process, which can be used to model both the renewal and non-renewal behaviors.

This paper is organized as follows. First we introduce the basic concept of
conditional intensity function in Sec.~\ref{sec2}. Then we show how to calculate
the conditional intensity function for electron emission in Sec.~\ref{sec3}. We
demonstrate the method by studying the electron emission in a dc-biased
single-channel QPC in Sec.~\ref{sec4}. The relation between the conditional
intensity function, WTD and joint WTD are also discussed in this
section. Finally, we summarize our results in Sec.~\ref{sec5}.

\section{conditional intensity function of a random point process}
\label{sec2}

Suppose one studies the electron emission process via an ideal single-electron
detector, then the emission process can be described by recording each emission
event in a time trace [See Fig.\ref{fig1}(a) for illustration]. This allows us
to represent emission events by random points in a line. This is quite similar
to the photon emission in quantum optics and neuronal spike emission in
neuroscience. Previous studies in these fields show that the emission process
can be described by the theory of random point process \cite{Snyder1991,
  D.J.Daley2003}. Moreover, one usually further assumes that two emission events
cannot occur exactly at the same time, \ie, there can only exist at most one
emission event in an arbitrary infinitesimal time interval $[t, t+dt)$. The
point process satisfied such assumption has been referred to as the regular
point process, which has been proved to be a valid assumption for typical
emission processes \cite{Kelley_1964, MANDEL1965, Macchi1975, Iwankiewicz1995}.

To characterize the statistics of the emission process, one usually needs a
proper probability distribution. In previous studies, the idle time distribution
$\Pi(t_s,t)$ has been introduced \cite{Albert2012}. In the theory of random
point process, it is also called the survivor function, which is written as
$S_1(t|t_s) = \Pi(t_s,t)$ \cite{D.J.Daley2003}. It gives the probability that no
electron is emitted in the time interval $[t_s,t]$, where $t_s$ can be treated
as the starting time of the detector.  Alternatively, one can also define the
emission probability density $p_1(t|t_s)$, which describes the probability of
the electron emission in the infinitesimal interval $[t, t+dt)$ since the
staring time $t_s$. The two distributions $S_1(t|t_s)$ and $p_1(t|t_s)$ are
equivalent, which can be related to each other as
\begin{equation}
  S_1(t|t_s) = 1 - \int^t_{t_s} d\tau p_1(\tau|t_s).
  \label{s2:eq1-1}
\end{equation}
By combing these two distributions, one can define a probability intensity
function as
\begin{equation}
  \lambda_1(t|t_s) = \frac{p_1(t|t_s)}{S_1(t|t_s)}.
  \label{s2:eq1}
\end{equation}
This function can be regarded as the conditional emission rate in the
infinitesimal interval $[t, t+dt)$, under the condition that no electron is
emitted in the time interval $[t_s, t]$.

The conditional emission rate $\lambda_1(t|t_s)$ essentially represents the
emission rate of the first electron since the starting time $t_s$. For a
history-independent process, it can play the role as the emission rate of the
whole process. This can be better illustrated by taking the stationary Poisson
point process as an example. The survivor function of the Poisson process can be
given as $S_1(t|t_s) = e^{-\lambda_0(t-t_s)}$, with $\lambda_0$ being the
emission rate. From Eqs.~\eqref{s2:eq1-1} and~\eqref{s2:eq1}, the corresponding
conditional probability can be given as $\lambda_1(t|t_s) = \lambda_0$.

The conditional emission rate can be generalized to incorporate the history
dependence of the emission process. Consider the emission rate of the $n$-th
($n \ge 2$) electron at time $t$, the corresponding history can be represented
by an ordered time sequence $\{ t_1, t_2, \dots, t_{n-1} \}$, which satisfies
\begin{equation}
  t_s < t_1 < t_2 < \dots < t_{n-1} < t
  \label{s2:eq1-2}.
\end{equation}
Here each $t_i$ ($i=1, 2, \dots, n-1$) represents the time instant of an
electron emission event that has already occurred since the starting time
$t_s$. To define the conditional emission rate in analogous to
Eq.~\eqref{s2:eq1}, one can generalize the emission probability density
$p_1(t|t_s)$ to the conditional emission probability density
$p_n(t|t_s, t_1, t_2, \dots, t_{n-1})$, which gives the emission probability of
the $n$-th electron at the time $t$ under the condition that there have been
$n-1$ electrons emitted previously in the infinitesimal interval
$[t_i, t_i + dt)$, respectively. The corresponding conditional survivor function
can be defined as
\begin{eqnarray}
  S_n(t| t_s,  t_1, t_2, \dots, t_{n-1} ) & = & 1 \nonumber\\
  &&\hspace{-3cm}{}- \int^t_{t_{n-1}} d\tau p_n( \tau| t_s,  t_1, t_2, \dots, t_{n-1}), 
  \label{s2:eq1-3}
\end{eqnarray}
which is formally analogous to Eq.~\eqref{s2:eq1-1}. Similarly, the conditional
emission rate for the $n$-th electron can be given as
\begin{equation}
  \lambda_n(t| t_s, t_1, t_2, \dots, t_{n-1}) = \frac{p_n(t| t_s, t_1, t_2, \dots, t_{n-1})}{S_n(t| t_s, t_1, t_2, \dots, t_{n-1})}.
  \label{s2:eq2}
\end{equation}
Due to the restriction given in Eq.~\eqref{s2:eq1-2}, the conditional emission
rates $\lambda_n(t| t_s, t_1, t_2, \dots, t_{n-1})$ with different $n$ can be
merged into one piecewise function $\lambda(t|\mathcal{H}_t)$, which has the
form
\begin{eqnarray}
  \lambda(t|\mathcal{H}_t) & = & \begin{cases}
    \lambda_1(t|t_s),          & t_s <  t  \le t_1 \\
    \lambda_2(t|t_s, t_1),  & t_1 <  t  \le t_2 \\
    \lambda_3(t|t_s, t_1, t_2), & t_2 <  t  \le t_3 \\
    \dots
  \end{cases}
  \label{s2:eq2-1},
\end{eqnarray}
where $\mathcal{H}_t = \{ t_1, t_2, \dots : t_s < t_1 < t_2 < \dots < t \}$
represents the history of the emission process up to the time $t$. It can be
treated as the instantaneous emission rate observed by an ideal single-electron
detector since the starting time $t_s$. In the theory of random process, it has
been referred to as the {\em conditional intensity function} \cite{Cox,
  D.J.Daley2003}. While it is less well-known in the context of mesoscopic
transport, it has been extensively used in other fields, such as the study of
neuronal spikes in neuroscience and random vibration analysis in civil
engineering \cite{Snyder1991, Iwankiewicz1995, D.J.Daley2003, Kass2014}.  With
the recent development of machine learning techniques, it can be extracted
effectively from the measured waiting times \cite{Du2016, Eichler2016}, leading
to potential applications in the data processing for real-time electron counting
techniques \cite{Gustavsson2009, Maisi2010, Kurzmann2018, Ranni2020,
  Brange2020}.

The conditional intensity function provides a time localized description of the
emission process, from which the temporal behavior of the emission process can
be understood intuitively. Moreover, it also contains the full information of
the emission process, from which various statistical quantities can be obtained.
In particular, both the WTD and joint WTD can be calculated from the conditional
intensity function.

The WTD $\mathcal{W}(t_s, t_e)$ is essentially a conditional probability density
\cite{Vyas1988, Brandes2008, Albert2011}. It gives the emission probability of
the second electron in the infinitesimal interval $[t_e, t_e+dt)$, under the
condition that the first electron has already been emitted at the time
$t_s$. Hence the WTD $\mathcal{W}(t_s, t_e)$ can be directly related to the
conditional emission probability density $p_2(t|t_s, t_1)$ as
\begin{eqnarray}
  \mathcal{W}(t_s, t_e) & = & p_2(t_e|t_s, t_s).
  \label{s5:eq10}
\end{eqnarray}
From Eqs.~\eqref{s2:eq1-3} and~\eqref{s2:eq2}, the conditional emission
probability $p_n(t|t_s, t_1, t_2, \dots, t_{n-1})$ can be obtained from the
conditional emission rates as
\begin{eqnarray}
  p_n(t|t_s, t_1, t_2, \dots, t_{n-1}) & = & \lambda_n(t|t_s, t_1, t_2, \dots, t_{n-1}) \nonumber\\
                                       &&\hspace{-2cm} \times e^{- \int^t_{t_{n-1}} d\tau \lambda_n(\tau|t_s, t_1, t_2, \dots, t_{n-1}) }.
  \label{s5:eq30}
\end{eqnarray}
So the WTD $\mathcal{W}(t_s, t_e)$ is direct related to the conditional emission
rate of the second electron $\lambda_2(t_e|t_s, t_1)$ with $t_1=t_s$.

Similarly, the joint WTD $\mathcal{W}_2(t_s, t_m, t_e)$ can also be obtained
from the conditional emission probability density as
\begin{eqnarray}
  \mathcal{W}_2(t_s, t_m, t_e) & = & p_2(t_m|t_s, t_s) p_3(t_e|t_s, t_s, t_m).
  \label{s5:eq20}
\end{eqnarray}
It gives the emission probability of the third electron in the infinitesimal
interval $[t_e, t_e+dt)$, under the condition that the first and second
electrons have been emitted at the time $t_s$ and $t_m$, respectively.  So the
joint WTD provides additional information on the conditional emission rate of
the third electron $\lambda_3(t|t_s, t_1, t_2)$ with $t_1=t_s$ and $t_2=t_m$. In
contrast, the conditional intensity function $\lambda(t|\mathcal{H}_t)$ from
Eq.~\eqref{s2:eq2-1} contains the conditional emission rate of all the $n$
electrons and hence provides a complete description of the whole emission
process.

\section{Electron emission as a determinantal point process}
\label{sec3}

For a general emission process, the conditional intensity function can be
calculated directly from the $n$-th order correlation functions
\cite{Kelley_1964}. Such calculation is usually rather involved. The calculation
can be greatly simplified in the non-interacting case, when all the correlation
functions can be expressed as determinants. In this case, the electron emission
can be modeled as a {\em determinantal point process} \cite{Macchi1975}. The
full information of such process can solely described by the first-order
correlation function.

For a single-channel QPC connected to two reservoirs $L$ and $R$ [See
Fig.~\ref{fig1}(a) for illustration], the first-order correlation function
$G(t,t')$ corresponding to the electron emission process can be cast into a
matrix form \cite{Haack2012}
\begin{eqnarray}
  G(t, t') & = & \left[ \begin{matrix} G_{RR}(t, t') & G_{RL}(t, t') \\
      G_{LR}(t, t') & G_{LL}(t, t') \\ \end{matrix} \right] \nonumber\\
           &=& \langle \Psi | \left[ \begin{matrix} \hat{b}^{\dagger}_{R}(t') & \hat{b}^{\dagger}_{L}(t') \end{matrix}
                                                                                \right] \left[ \begin{matrix} \hat{b}_{R}(t) \\ \hat{b}_{L}(t) \\ \end{matrix} \right] | \Psi \rangle \nonumber\\
           &&\mbox{}- \langle F | \left[ \begin{matrix} \hat{a}^{\dagger}_{R}(t') & \hat{a}^{\dagger}_{L}(t') \end{matrix}
                                                                                       \right] \left[ \begin{matrix} \hat{a}_{R}(t) \\ \hat{a}_{L}(t) \\ \end{matrix} \right] | F \rangle,
  \label{s3:eq10}  
\end{eqnarray}
where $| \Psi \rangle$ represents the many-body state of the emitted electrons,
while $| F \rangle$ represents the undisturbed Fermi sea. The incoming and
outgoing electrons are represented by Fermion operators $\hat{a}_{\eta}(t)$ and
$\hat{b}_{\eta}(t)$ ($\eta = L, R$), respectively. These operators can be
related to each other via the scattering matrix as
\begin{equation}
  \left[ \begin{matrix}
       \hat{b}_{R}(t)\\
       \hat{b}_{L}(t)\\
    \end{matrix}\right] = \left[ \begin{matrix}
        \sqrt{1-D} & i\sqrt{D} e^{-i\phi(t)} \\
        i\sqrt{D} e^{i\phi(t)}  & \sqrt{1-D} \\
    \end{matrix}\right] \left[ \begin{matrix}
       \hat{a}_{R}(t)\\
       \hat{a}_{L}(t)\\
    \end{matrix}\right],
  \label{s3:eq20}
\end{equation}
with $D$ representing the transparency of the QPC and $\phi(t)$ being the
forward scattering phase.

When the QPC is driven by a constant bias voltage $V$, the forward scattering
phase can be expressed as $\phi(t) = eVt/\hbar$. In this case, the first-order
correlation function can be decomposed in terms of Martin-Landauer wave packets
as
\begin{widetext}
  \begin{eqnarray}
    \hspace{-0.5cm}\left[ \begin{matrix} G_{RR}(t, t') & G_{RL}(t, t') \\
        G_{LR}(t, t') & G_{LL}(t, t') \\ \end{matrix} \right] & = & \sum_{l = 0, \pm 1, \pm 2, \dots} \left[ \begin{matrix}
        D \psi_e(t-lT) \psi^\ast_e(t'-lT) & -i\sqrt{D(1-D)} \psi_h(t-lT) \psi^\ast_e(t'-lT) \\
        i\sqrt{D(1-D)} \psi_e(t-lT) \psi^\ast_h(t'-lT) & D \psi_h(t-lT) \psi^\ast_h(t'-lT) \\
      \end{matrix}\right]
    \label{s3:eq30},
  \end{eqnarray}
\end{widetext}
with $T = h/eV$ representing the repetition period and
\begin{eqnarray}
  \psi_e(t) & = & \frac{1}{\sqrt{T}} e^{-i \pi t/T} \frac{\sin(\pi t/T)}{\pi t/T}, \nonumber\\
  \psi_h(t) & = & \frac{1}{\sqrt{T}} e^{i \pi t/T} \frac{\sin(\pi t/T)}{\pi t/T},
              \label{s3:eq31}
\end{eqnarray}
representing the Martin-Landauer wave packets corresponding to the emitted
electrons and holes, respectively.


From the above expression, one can see that the four components of the
first-order correlation function $G_{\eta\eta'}(t, t')$ have different physical
meanings: $G_{RR}(t, t')$ [$G_{LL}(t, t')$] describes the emission of electrons
[holes] into the right [left] reservoir. In contrast, $G_{LR}(t, t')$
[$G_{RL}(t, t')$] corresponds to the emission of electron-hole pairs: While the
electron [hole] component is emitted to the left [right] reservoir, the hole
[electron] component is reflected back to the right [left] reservoir.

In this paper, we focus on the electron emission into the right reservoirs,
which corresponds to the component $G_{RR}(t, t')$. To calculate the conditional
intensity function by using Eqs.~\eqref{s2:eq1-3} and~\eqref{s2:eq2}, one needs
information of the survivor function $S_n(t| t_s, t_1, t_2, \dots, t_{n-1} )$ or
the conditional emission probability $p_n(t| t_s, t_1, t_2, \dots, t_{n-1}
)$. In a previous work, Macchi has shown that they can be extracted from
$G_{RR}(t, t')$ via the following procedure \cite{Macchi1975}:
\begin{itemize}
\item Solve the eigenvalue equation:
  \begin{equation}
    \int^t_{t_s} d\tau' G_{RR}(\tau, \tau') \varphi_{\alpha}(\tau') = \nu_{\alpha} \varphi_{\alpha}(\tau),
    \label{s3:eq40}
  \end{equation}
  with $\alpha = 1, 2, \dots$ being the index of the eigenvalues and
  eigenfunctions. The eigenvalue $\nu_{\alpha}$ satisfies
  $ 0 \le \nu_{\alpha} \le 1$, while the eigenfunctions $\varphi_{\alpha}(\tau)$
  form an orthonormal basis within the time interval $[t_s, t]$, \ie,
  \begin{equation}
    \int^{t}_{t_s} d\tau \varphi^{\ast}_{\alpha}(\tau) \varphi_{\alpha'}(\tau) = \delta_{\alpha, \alpha'}.
    \label{s3:eq50}
  \end{equation}
  From the eigenvalues and eigenfunctions, one can define an auxiliary function
  as
  \begin{equation}
    C(t, t') = \sum_{\alpha=1}^{+\infty} \frac{\nu_\alpha}{1-\nu_\alpha}
    \varphi_{\alpha}(t) \varphi^{\ast}_{\alpha}(t').
    \label{s3:eq60}
  \end{equation}
\item Define the exclusive density function $\pi_n(t_1, t_2, \dots, t_n|t_s, t)$
  as
  \begin{eqnarray}
    \pi_n(t_1, t_2, \dots, t_n|t_s, t) & = & p_1(t_1|t_s) \nonumber\\
    &&\hspace{-2.5cm} \times p_2(t_2|t_s, t_1) \cdots p_n(t_n|t_s, t_1, \dots, t_{n-1})
                                             \nonumber\\
    &&\hspace{-2.5cm} \times S_{n+1}(t|t_s, t_1, t_2, \dots, t_n),
  \end{eqnarray}
  with $n \ge 1$. This function can be calculated from $C(t, t')$ and
  $\nu_\alpha$ as
  \begin{eqnarray}
    && \pi_n(t_1, t_2, \dots, t_n|t_s,t) =  \Big[ \prod_{\alpha=1}^N (1 - \nu_\alpha) \Big] \nonumber\\
    && \times \begin{vmatrix}
      C(t_1, t_1) & C(t_1, t_2) & \dots & C(t_1, t_n) \\
      C(t_2, t_1) & C(t_2, t_2) & \dots & C(t_2, t_n) \\
      \hdotsfor{1}                                    \\
      C(t_n, t_1) & C(t_n, t_2) & \dots & C(t_n, t_n)
      \label{s3:eq70}
    \end{vmatrix}.
  \end{eqnarray}
\item The corresponding idle time distribution $\Pi(t_s,t)$, or equivalently the
  survivor function $S_1(t|t_s)$, can be solely determined by the $\nu_\alpha$
  as
  \begin{equation}
    \Pi(t_s, t) = S_1(t|t_s) = \prod_{\alpha=1}^N (1 - \nu_\alpha).
    \label{s3:eq80}
  \end{equation}
  Equation~\eqref{s3:eq70} and~\eqref{s3:eq80} can be used to extract
  $S_n(t| t_s, t_1, t_2, \dots, t_{n-1} )$ and
  $p_n(t| t_s, t_1, t_2, \dots, t_{n-1} )$.
\end{itemize}
Then all the conditional emission rates
$\lambda_n(t|t_s, t_1, t_2, \dots, t_{n-1})$ can be calculated by using
Eqs.~\eqref{s2:eq1-1},~\eqref{s2:eq1},~\eqref{s2:eq1-3} and~\eqref{s2:eq2}. For
example, the conditional emission rate for the first and second electrons can be
written as
\begin{eqnarray}
  \lambda_1(t|t_s) & = & C(t, t), \\
  \lambda_2(t|t_s, t_1) & = & \frac{C(t_1, t_1)C(t, t) - \left| C(t_1, t) \right|^2}{C(t, t)}.
                                  \label{s3:eq82}
\end{eqnarray}
This provides an efficient numerical methods to evaluate the conditional intensity function given in
Eq.~\eqref{s2:eq2-1}.

It is worth noting that the eigenvalue $\nu_\alpha$ obtained from
Eq.~\eqref{s3:eq40} can be treated as the emission probability of the
$\alpha$-th electron emitted in the time interval $[t_s, t]$. This can be seen
from the time-dependent full counting statistics (FCS). The corresponding
momentum generating function of the FCS can be given as \cite{Macchi1975}
\begin{equation}
  \Phi(\chi) = \prod_{\alpha=1}^{+\infty} \left( 1 - \nu_\alpha + e^{i \chi} \nu_\alpha \right).
  \label{s3:eq90}
\end{equation}
This corresponds to a generalized binomial statistics, which indicates that
within a finite time interval $[t_s, t]$, the $\alpha$-th electron attempts to
emit with a success probability $\nu_\alpha$. Note that the probability
$\nu_\alpha$ is not a constant, but typically time-dependent. Hence the
corresponding emission events should not be considered as independent Bernoulli
trials, but are time-correlated.

General speaking, all the results should also depends on the starting time
$t_s$. This is is irrelevant for the dc-biased QPC, as the emission process is
stationary. Without loss of generality, in the following discussion we always
choose $t_s=0.0$. In this case, the WTD and joint WTD from Eqs.~\eqref{s5:eq10}
and~\eqref{s5:eq20} can be written as
\begin{eqnarray}
  \mathcal{W}(\tau) & = & \mathcal{W}(t_s, t_s + \tau), \label{s5:eq40-1}\\
  \mathcal{W}_2(\tau_1, \tau_2) & = & \mathcal{W}_2(t_s, t_s + \tau_1, t_s + \tau_1 + \tau_2). \label{s5:eq40-2}
\end{eqnarray}

\section{Emission rates for QPC}
\label{sec4}

From the discussion in the above section, one can see that, for a dc-biased QPC,
the emitted electrons can be represented by a sequence of Martin-Landauer wave
packets with repetition period $T$, which can transmit across the QPC with a
finite probability $D$. The average emission rate of the electron can then be
given as $\bar{\lambda} = D/T$. Here the repetition period $T=h/eV$ is decided
by the bias voltage, while the probability $D$ is just equal to the QPC
transparency. The two parameters $T$ and $D$ have different impacts on the
emission process.

\begin{figure}
  \centering
  \includegraphics[width=8.0cm]{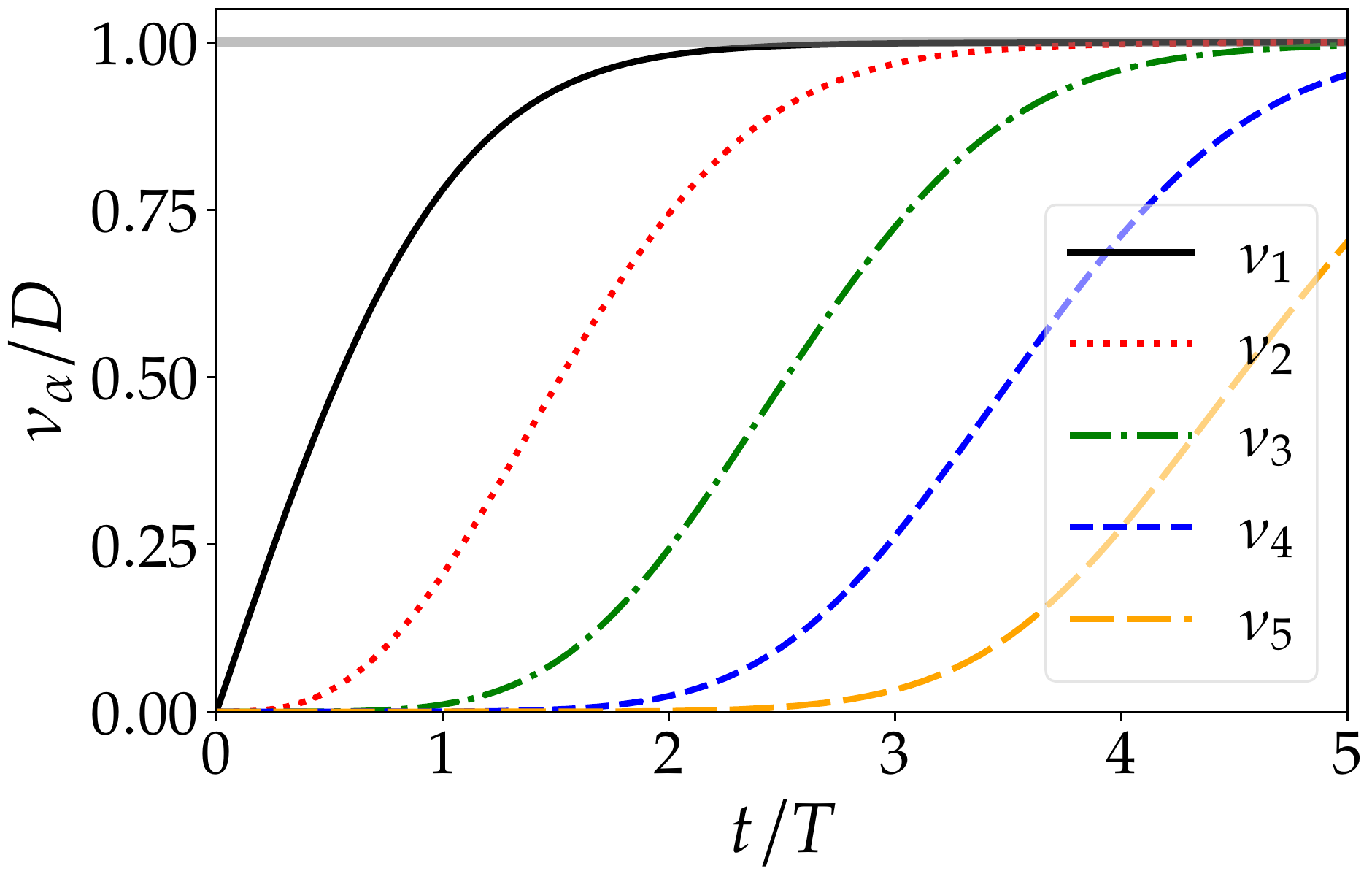}
  \caption{ Emission probabilities $\nu_\alpha$ as functions of time $t$. The
    horizontal line is restricted to the QPC transparency $D$.}
  \label{fig2}
\end{figure}

To see this, let us first concentrate on the emission probabilities
$\nu_\alpha$. From Eqs.~\eqref{s3:eq30},~\eqref{s3:eq31} and~\eqref{s3:eq40},
one can see that the time-dependence of $\nu_\alpha$ is mainly decided by the
repetition period $T$. Indeed, as the time $t$ increases, all the emission
probabilities increase monotonically and saturate on timescales comparable to
$T$. This is illustrated in Fig.~\ref{fig2}. In contrast, the QPC transparency
$D$ merely plays the role of a scale factor, which only restricts the saturation
value of the emission probabilities to $D$ [illustrated by the grey line in
Fig.~\ref{fig2}].

\begin{figure}
  \centering
  \includegraphics[width=8.0cm]{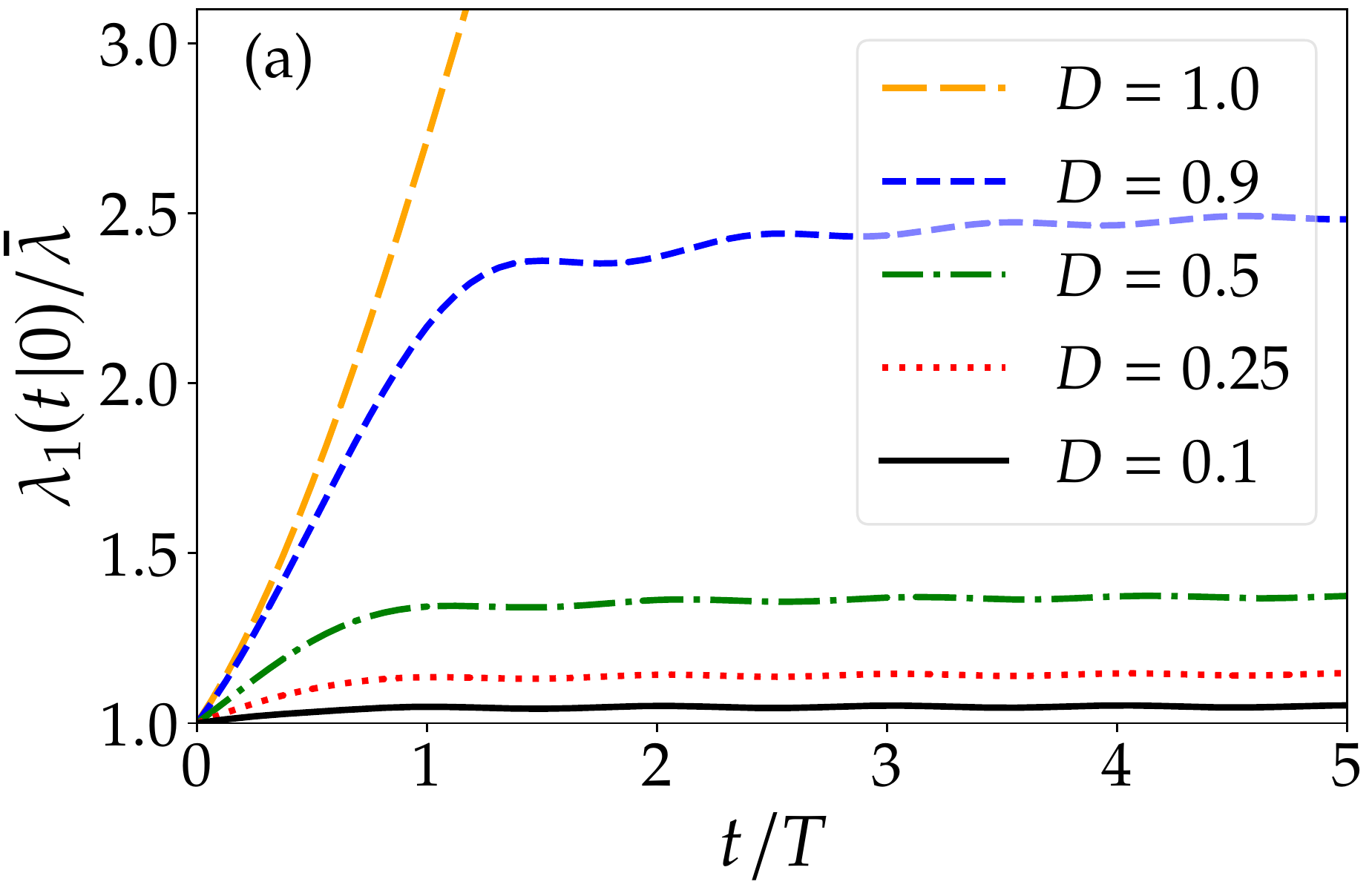}
  \includegraphics[width=8.0cm]{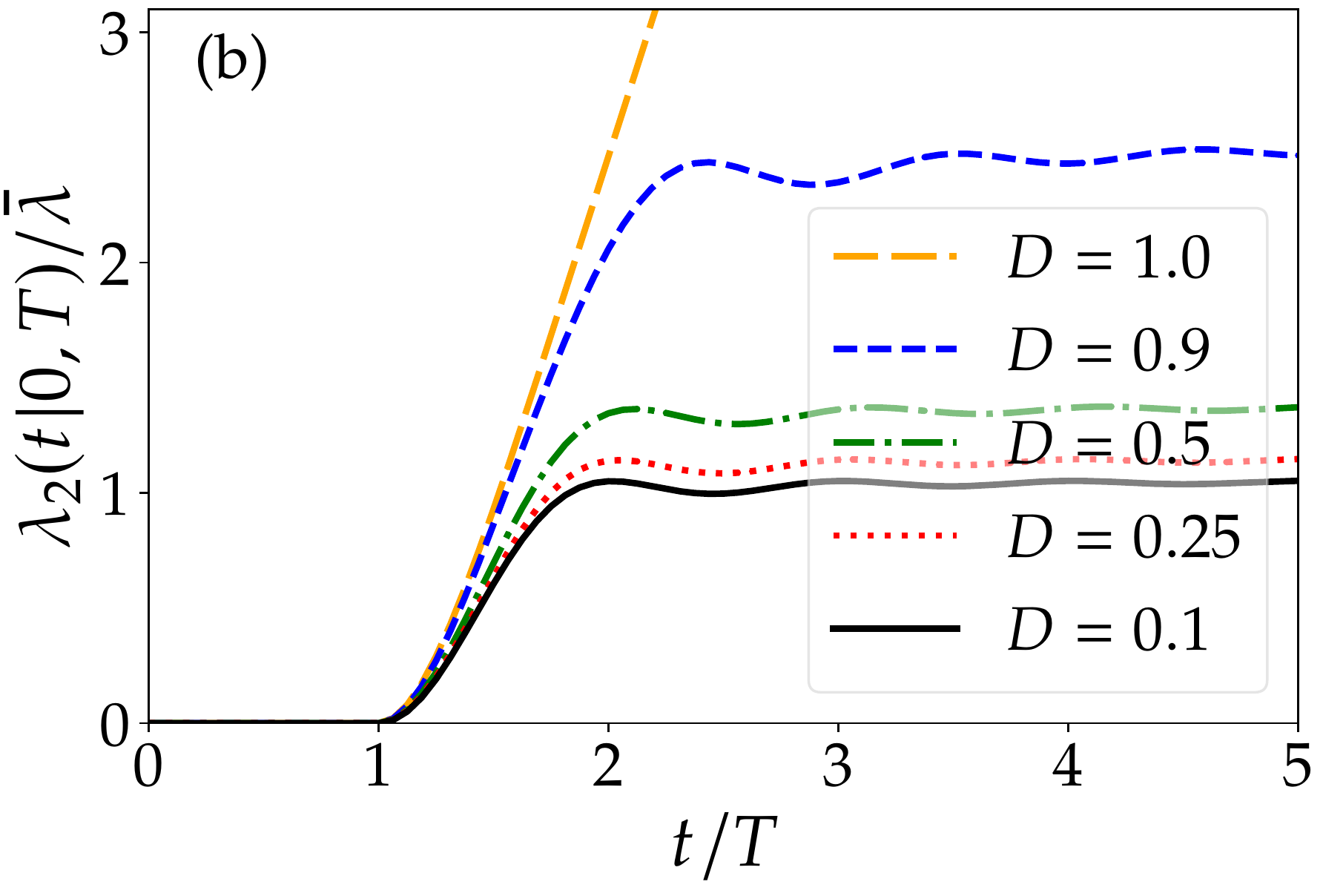}
  \caption{ (a) The conditional emission rate of the first electron
    $\lambda_1(t|0)$ (b) The conditional emission rate of the second electron
    $\lambda_2(t|0, T)$. In both figures, curves with different curves with
    different colors and line types correspond to different QPC transparency
    $D$.}
  \label{fig3}
\end{figure}

Although the impact of $D$ on the emission probabilities is trivial, it has a
much pronounced impact on the emission rates. This can be seen from the
conditional emission rate $\lambda_1(t|0)$ of the first electron, which is
illustrated in Fig.~\ref{fig3}(a). In the figure, curves with different colors
and line types correspond to $\lambda_1(t|0)$ with different transparency
$D$. At the time $t/T=0.0$, $\lambda_1(t|0)$ is equal to the average emission
rate $\bar{\lambda}$. As $t$ increases, it start to increase, which shows
different behaviors for different $D$: For $D < 1.0$, $\lambda_1(t|0)$ saturates
on the timescale of $T$ and undergoes a weak oscillation around the saturation
value as $t$ further increases. This can be seen from the black solid, red
dotted, green dash-dotted and blue dashed curves, corresponding to $D=0.1$,
$0.25$, $0.5$ and $0.9$, respectively. In contrast, $\lambda_1(t|0)$ does not
saturate at all for $D=1.0$. It increases almost linearly as a function of $t$,
which is illustrated by the orange long dashed curve in the figure.

Similar behaviors can also be found for other conditional emission rates. To
demonstrate this, we plot the conditional emission rate of the second electron
$\lambda_2(t|0, T)$ in Fig.~\ref{fig3}(b). Curves with different colors and line
types correspond to $\lambda_2(t|0, T)$ with different transparency $D$. From
the figure, one always finds $\lambda_2(t|0, T) = 0.0$ for $t=T$, indicating
that the emission of the second electron is hindered by the first electron
emitted at the time instant $t_1=T$. For $D < 1.0$, $\lambda_2(t|0, T)$
increases as a function of $t$ and saturates before $t/T = 2.0$. Then it
undergoes a weak oscillation around the saturation value as $t$ further
increases. These behaviors are illustrated by the black solid, red dotted, green
dash-dotted and blue dashed curves, corresponding to $D=0.1$, $0.25$, $0.5$ and
$0.9$, respectively. For $D=1.0$, $\lambda_2(t|0, T)$ increases almost linearly
as a function of $t$ and does not saturate at all. This can be seen from the
orange long dashed curve. These results indicates that the rising time of the
conditional emission rates are decided solely by the repetition period $T$.

The conditional intensity function $\lambda(t|\mathcal{H}_t)$ can be obtained by
combining all the conditional emission rates following Eq.~\eqref{s2:eq2-1}. To
do this, one needs to choose a proper set of $t_i$, which represents the time
instants of emission events in the history $\mathcal{H}_t$ [see the definition
of $\mathcal{H}_t$ below Eq.~\eqref{s2:eq2-1}]. Due to the probabilistic nature
of the electron emission, $t_i$ are essentially random parameters. Their waiting
times $\tau=t_i-t_{i-1}$ follow the WTD $\mathcal{W}(\tau)$, whose mean value is
equal to the average waiting time $1/\bar{\lambda}$. In order to simplify the
discussion, we choose $t_i = i/\bar{\lambda}$ in the following calculation.

\begin{figure}
  \centering
  \includegraphics[width=8.0cm]{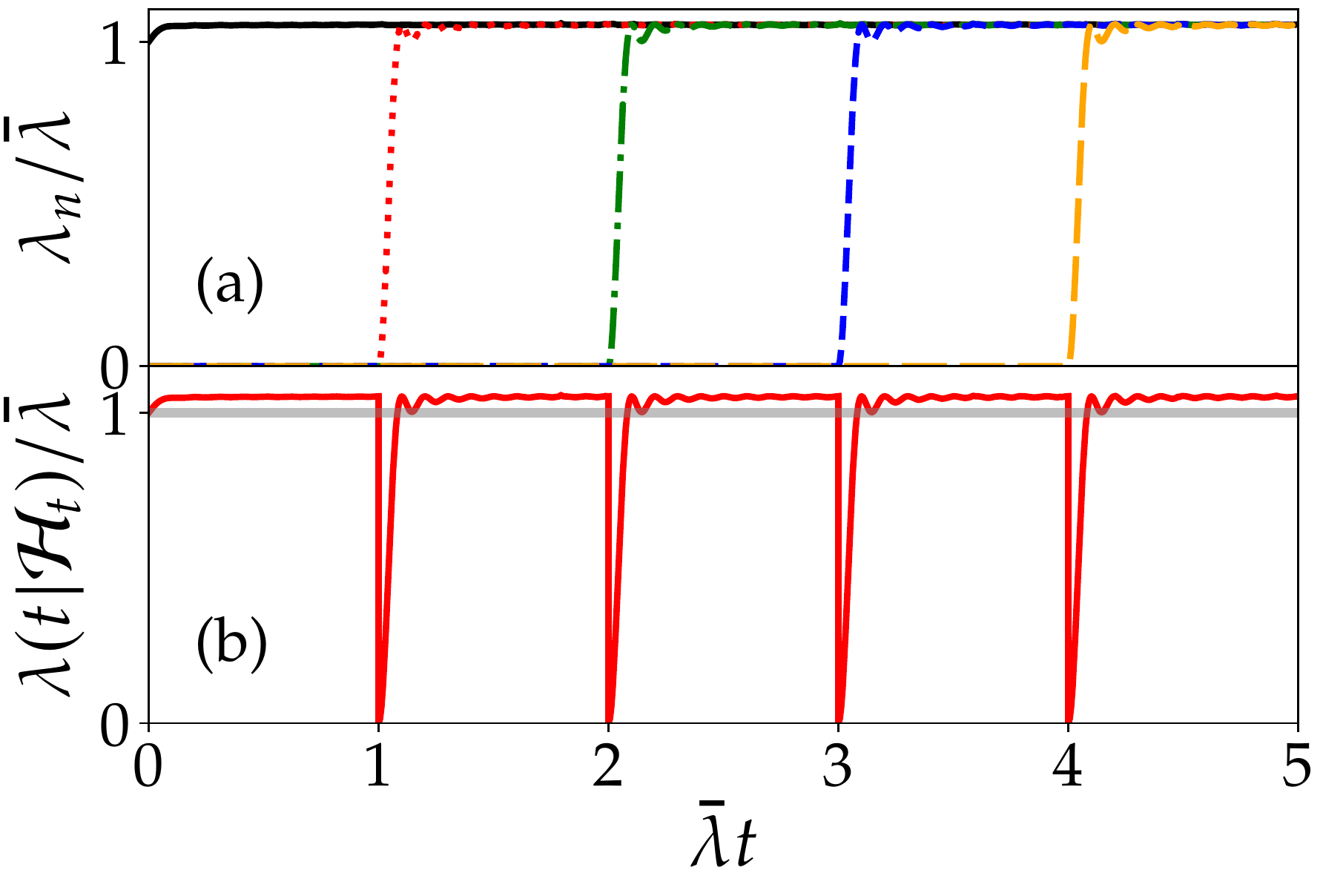}
  \caption{ (a) Conditional emission rates
    $\lambda_n(t|0, t_1, t_2, \dots, t_{n-1})$ for the first five electrons. The
    black solid, red dotted, green dash-dotted, blue dashed, and orange
    long-dashed curves correspond to $n=1, 2, 3, 4$ and $5$, respectively. (b)
    The conditional intensity function $\lambda(t|\mathcal{H}_t)$ obtained from
    the conditional emission rates via Eq.~\eqref{s2:eq2-1}. The average
    emission rate $\bar{\lambda}$ is plotted by the horizontal grey line. All
    the emission rates and $t$ are rescaled according to the average emission
    rate $\bar{\lambda}$.}
  \label{fig4}
\end{figure}

In the case of low transparency, the rising time of the conditional emission
rates is much shorter than the average waiting time. This case is demonstrated
in Fig.~\ref{fig4}, corresponding to $D=0.1$. In Fig.~\ref{fig4}(a), we plot the
conditional emission rates for the first five emitted electrons with different
curves. Due to the short rising time, one can see that the emission rate can
exhibit a step-like increase around the time $t=t_i$. The oscillation around the
saturation value is also rather small. This makes the corresponding conditional
intensity function can be well-approximated by the average emission rate
$\bar{\lambda}$ in most regions, as illustrated in Fig.~\ref{fig4}(b). The
approximation is only invalid at the vicinity of $t_i$, where the conditional
intensity function exhibits sharp dips. These dips is not crucial for long-time
properties, such as dc shot noise. This makes the electron emission can be
treated approximated as a simple Poisson process on long time scales, which can
be characterized via a constant emission rate.

\begin{figure}
  \centering
  \includegraphics[width=8.0cm]{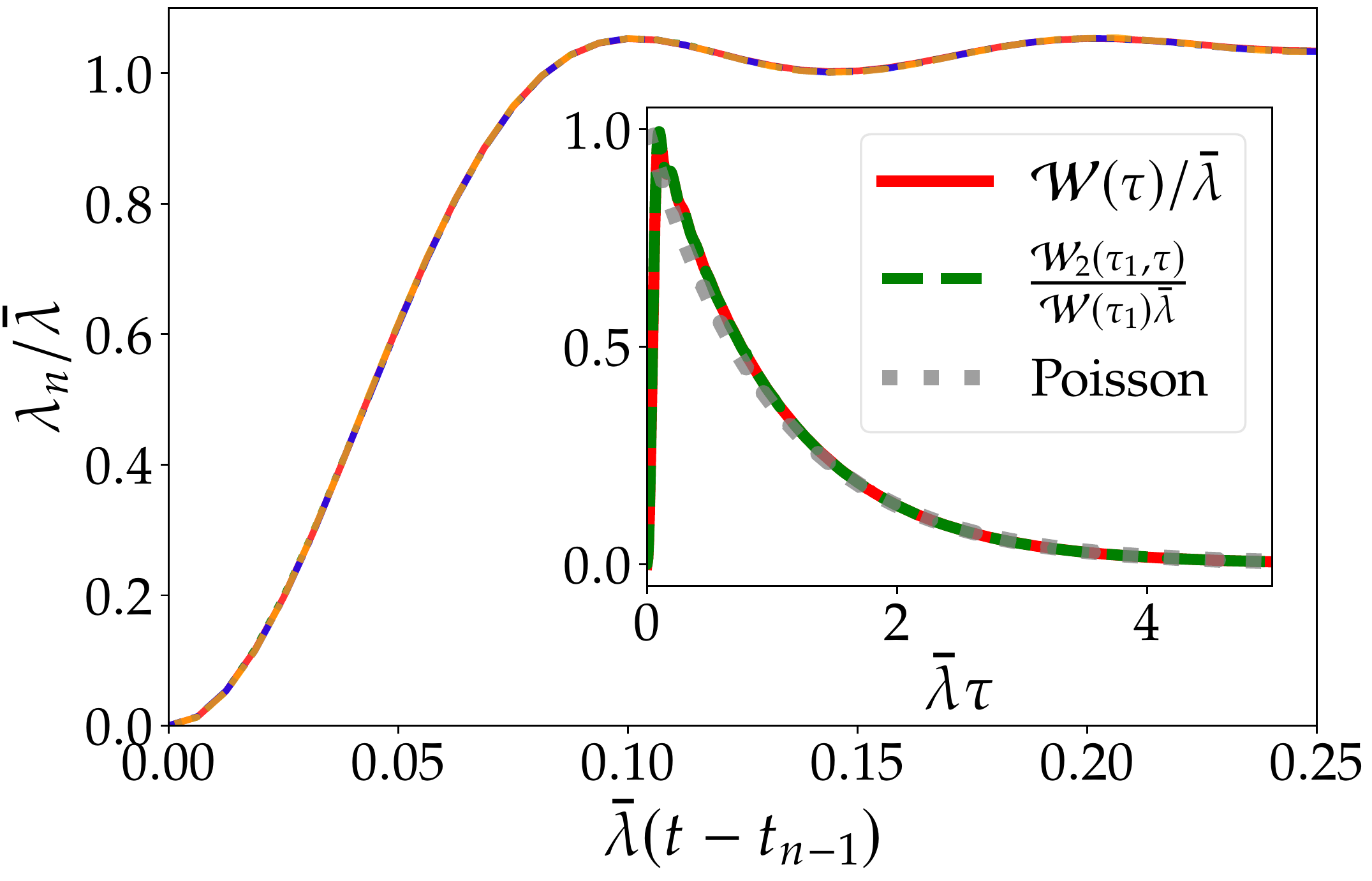}
  \caption{ Main panel: Conditional emission rates
    $\lambda_n(t|0, t_1, t_2, \dots, t_{n-1})$ as a function of the waiting time
    $t-t_{n-1}$. The red solid, green dotted, blue dash-dotted, and orange
    dashed curves correspond to $n=2, 3, 4$ and $5$, respectively. Note that all
    the curves are highly overlapped.  Inset: Rescaled WTD
    $\mathcal{W}(\tau)/\bar{\lambda}$ and joint WTD
    $\mathcal{W}_2(\tau_1, \tau)/[\mathcal{W}(\tau_1)\bar{\lambda}]$ as
    functions of $\bar{\lambda}\tau$. The grey dotted curve in the inset
    corresponds to the WTD of the simple Poisson process.}
  \label{fig5}
\end{figure}

As long as short-time behaviors are concerned, the Poisson approximation breaks
down. In this case, one has to characterize the emission rate via the
conditional intensity function $\lambda(t|\mathcal{H}_t)$, which generally has a
complicated history-dependence. However, in the case of low transparency, we
find that the conditional emission rate
$\lambda_n(t|0, t_1, t_2, \dots, t_{n-1})$ do not depend on the whole history
$\mathcal{H}_t$, but is only sensitive to the time instant of the emission of
the previous electron. As a consequence, all the conditional emission rates have
a similar profile as a function of the waiting time $t - t_{n-1}$. This is
illustrated in the main panel of Fig.~\ref{fig5}. In the figure, we plot the
conditional emission rate $\lambda_n(t|0, t_1, t_2, \dots, t_{n-1})$ as the
function of $t-t_{n-1}$, where the red solid, green dotted, blue dash-dotted,
and orange dashed curves correspond to $n=2, 3, 4$ and $5$, respectively. One
can see that all the curves coincide with each other. This indicates that they
can be expressed as
$\lambda_n(t|0, t_1, t_2, \dots, t_{n-1}) = \lambda_r(t-t_{n-1})$, which
essentially corresponds to a stationary renewal process.

The renewal behavior can also be seen from the joint WTD analysis. For a renewal
process, one expects that
$\mathcal{W}_2(\tau_1, \tau_2) = \mathcal{W}(\tau_1) \mathcal{W}(\tau_2)$,
indicating that there are no correlation between waiting times. To demonstrate
this, we plot the rescaled WTD $\mathcal{W}(\tau)/\bar{\lambda}$ [calculated
from Eqs.~\eqref{s5:eq10},~\eqref{s5:eq30} and~\eqref{s5:eq40-1}] by the red
solid curve in the inset of Fig.~\ref{fig5}.  The corresponding rescaled joint
WTD $\mathcal{W}_2(\tau_1, \tau)/[\mathcal{W}(\tau_1)\bar{\lambda}]$ is plotted
with the green dotted curves, where we have chosen $\tau_1 = 1/\bar{\lambda}$.
One can see that two curves agree quite well, indicating the absence of the
correlation between waiting times. Moreover, one can see that both WTDs can be
well-approximated by an exponential distribution (grey curve) for large $\tau$,
indicating that the process can be treated as a simple Poisson process on long
time scales.

\begin{figure}
  \centering
  \includegraphics[width=8.0cm]{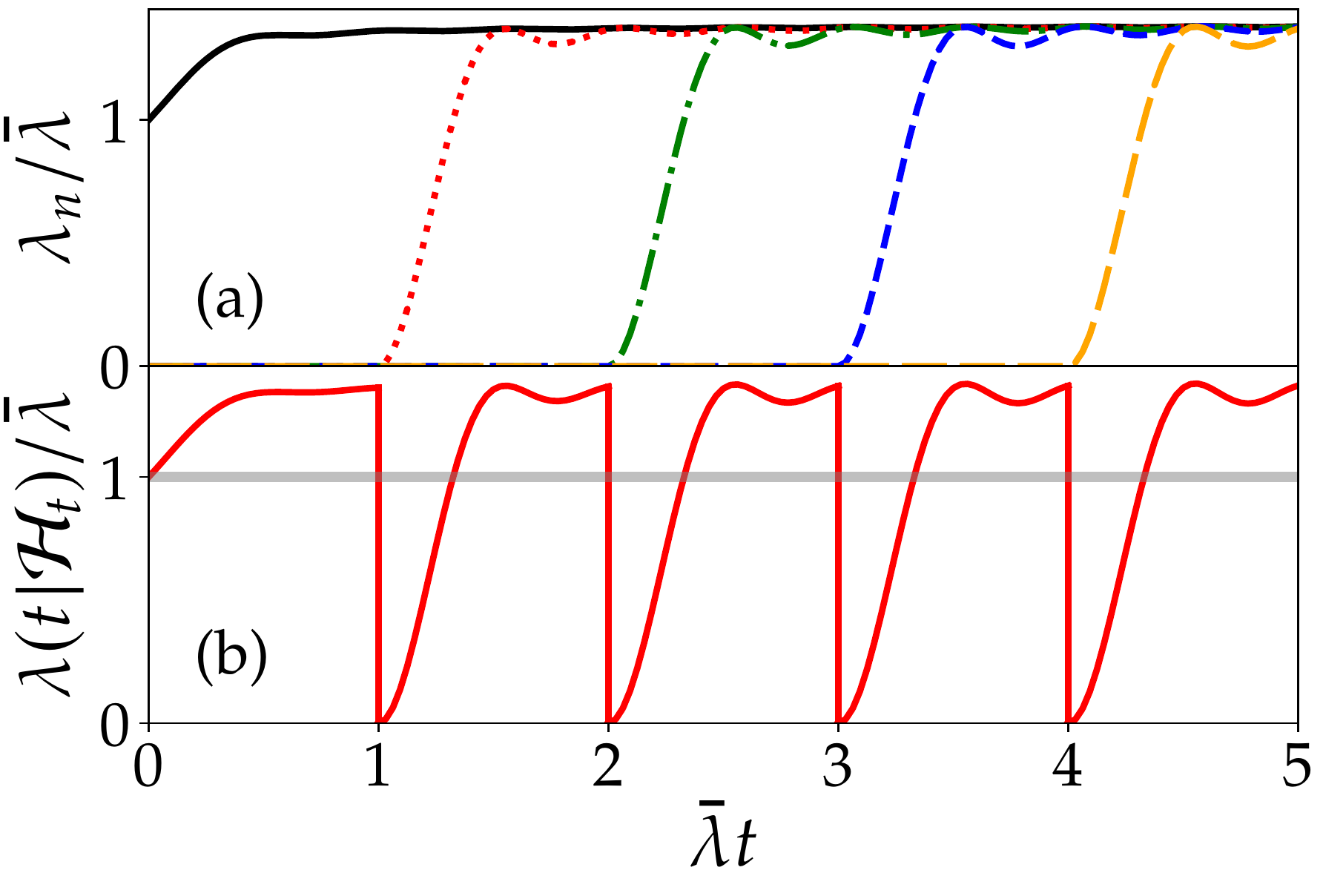}
  \caption{ The same as Fig.~\ref{fig4}, but with $D=0.5$. }
  \label{fig6}
\end{figure}

\begin{figure}
  \centering
  \includegraphics[width=8.0cm]{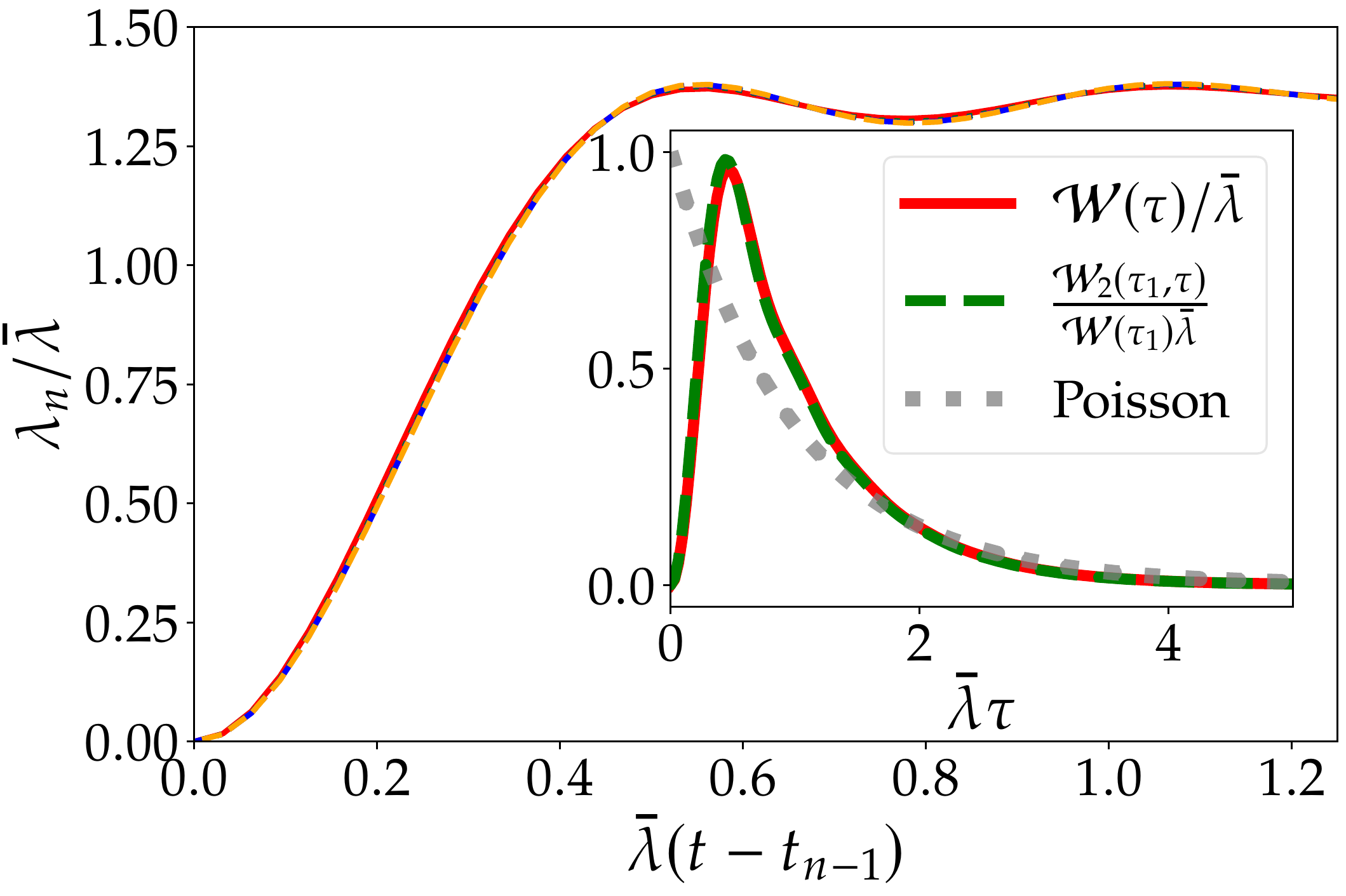}
  \caption{ The same as Fig.~\ref{fig5}, but with $D=0.5$. }
  \label{fig7}
\end{figure}

For the QPC with modest transparency, the rising time of the conditional
emission rates can be comparable to the average waiting time. This case is
demonstrated in Fig.~\ref{fig6}(a), corresponding to $D=0.5$. Due to the long
rising time, the dips evolves into wide valleys. Moreover, the saturation value
is also much larger then the average emission rate. One can also see a
pronounced oscillation in the saturation region. These features can be seen from
Fig.~\ref{fig6}(b). All these features indicates that the emission process
cannot be approximated as a stationary Poisson process even on long time
scales. However, we find that all the conditional emission rates
$\lambda_n(t|0, t_1, t_2, \dots, t_{n-1})$ as a function of the waiting time
$t-t_{n-1}$ still have a similar profile, which is illustrated in the main panel
of Fig.~\ref{fig7}. This indicates that the emission process can still be
treated as a renewal process.

The renewal behavior can also be seen from the WTD analysis, which is
illustrated in the inset of Fig.~\ref{fig7}. As the two rescaled WTDs coincides,
the join WTD can still be approximated as
$\mathcal{W}_2(\tau_1, \tau) = \mathcal{W}(\tau_1) \mathcal{W}(\tau)$,
indicating the absence of correlation between waiting times. Note that the
profile of the WTDs are significantly different from an exponential
distribution, indicating a departure from the Poisson approximation.

\begin{figure}
  \centering
  \includegraphics[width=8.0cm]{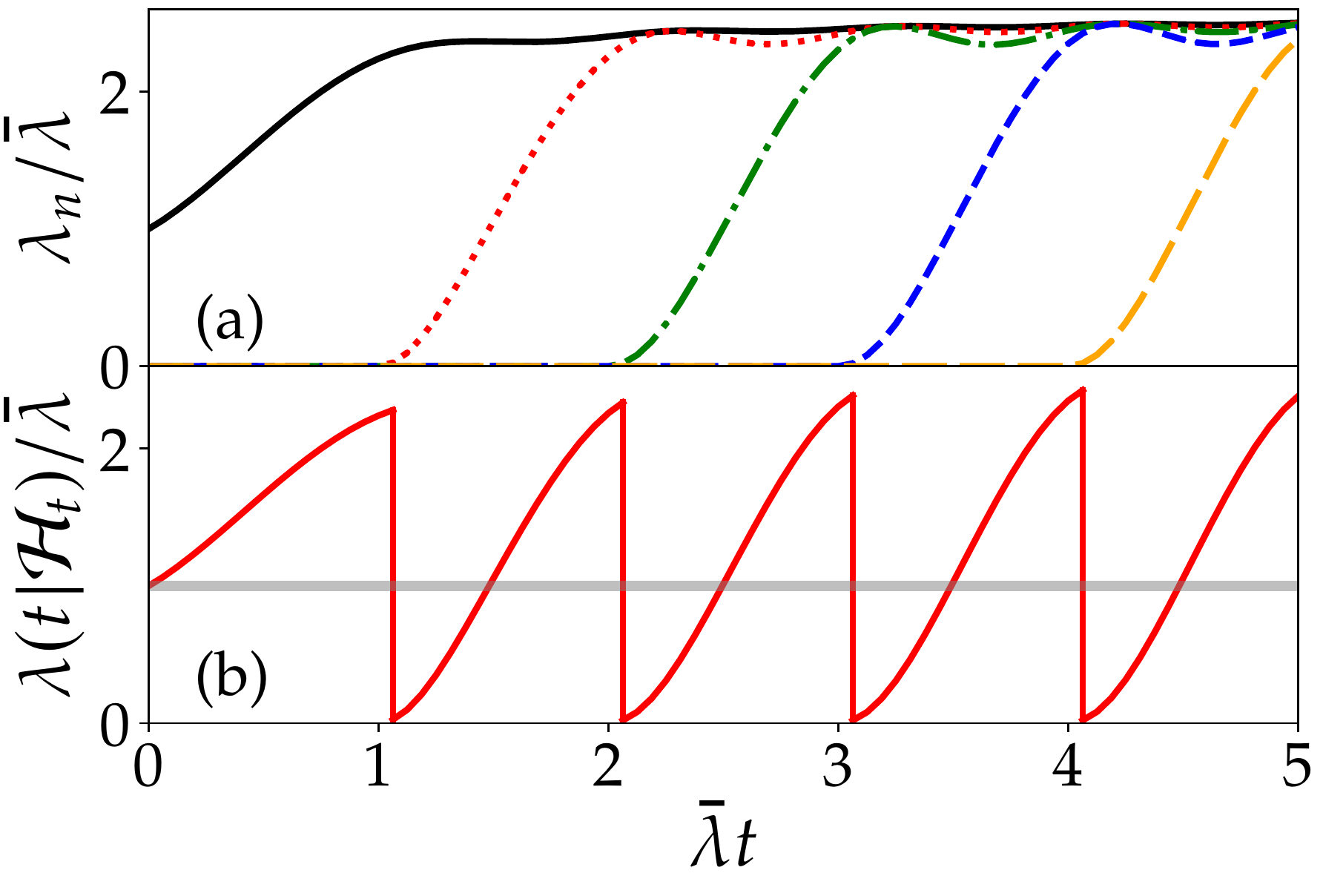}
  \caption{ The same as Fig.~\ref{fig4}, but with $D=0.9$. }
  \label{fig8}
\end{figure}

\begin{figure}
  \centering
  \includegraphics[width=8.0cm]{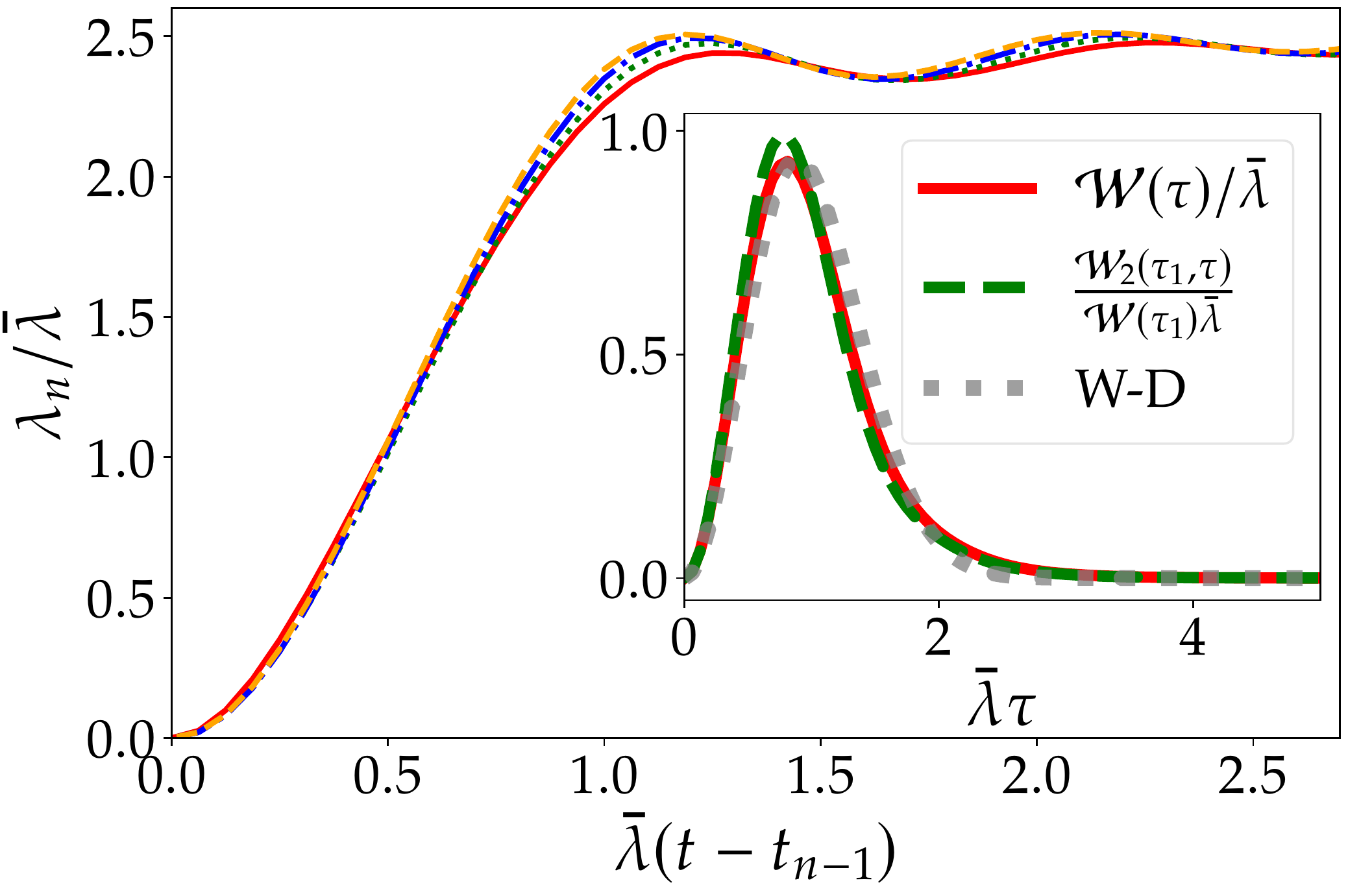}
  \caption{ The same as Fig.~\ref{fig5}, but with $D=0.9$. The grey dotted curve
    in the inset corresponds to the Wigner-Dyson distribution.}
  \label{fig9}
\end{figure}

As the QPC is further opened up, the rising time of the conditional emission
rates can be longer than the average waiting time. This case is demonstrated in
Fig.~\ref{fig8}(a), corresponding to $D=0.9$. In this case, the conditional
intensity function exhibits a saw-tooth behavior in the time domain, indicating
the presence of strong correlations. This makes the renewal approximation breaks
down: The conditional emission rates for different electrons can exhibit
different time dependence, which cannot be described by a universal rate
function $\lambda_r(t-t_{n-1})$. This is demonstrated in the main panel of
Fig.~\ref{fig9}. Accordingly, the relation
$\mathcal{W}_2(\tau_1, \tau) = \mathcal{W}(\tau_1) \mathcal{W}(\tau)$ does not
hold, which is plotted in the inset of Fig.~\ref{fig9}. Note that the WTD in
this case is quite close to the Wigner-Dyson distribution, which is plotted by
the grey dotted curve in the inset.

\begin{figure}
  \centering
  \includegraphics[width=8.0cm]{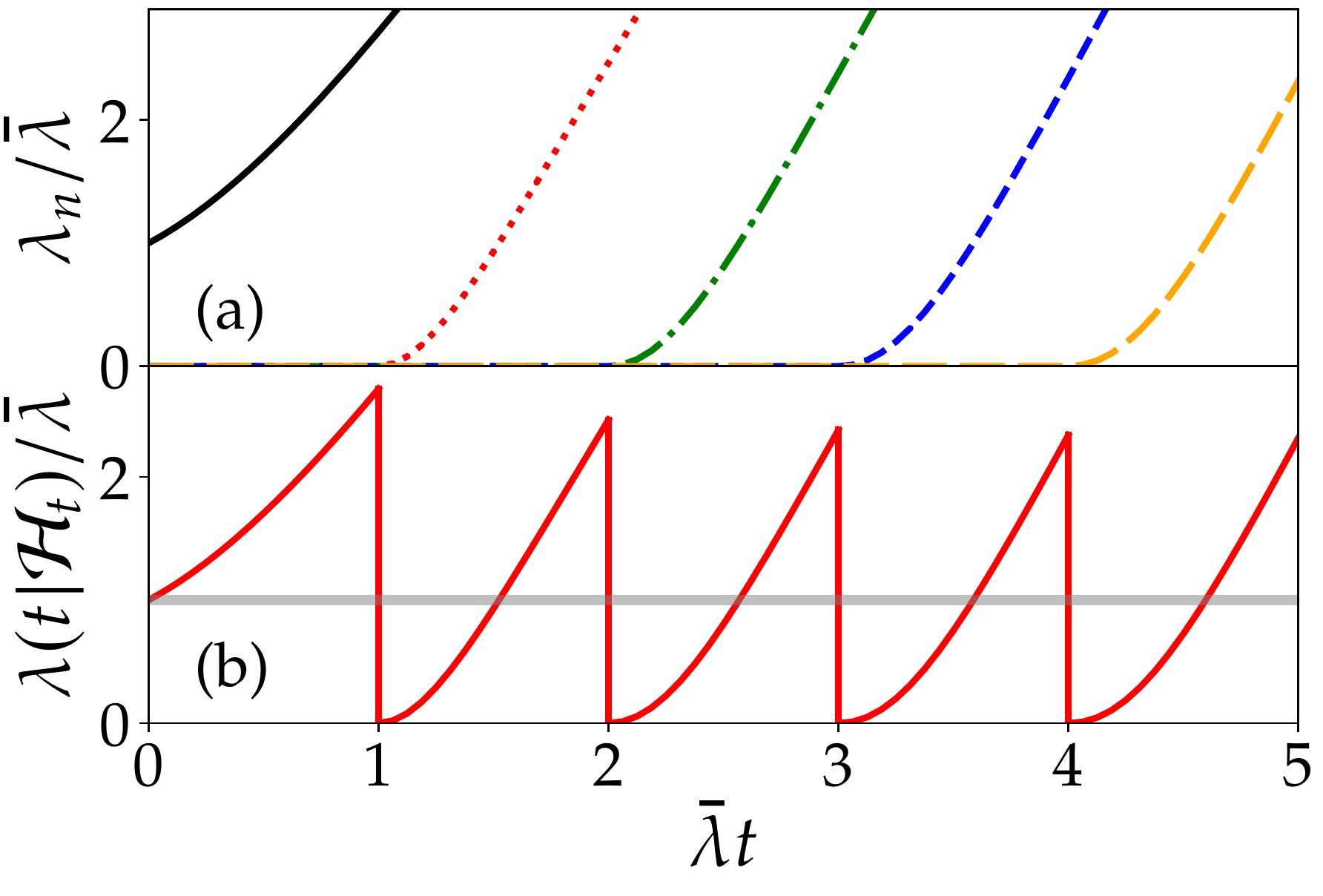}
  \caption{The same as Fig.~\ref{fig4}, but with $D=1.0$.}
  \label{fig10}
\end{figure}

\begin{figure}
  \centering
  \includegraphics[width=8.0cm]{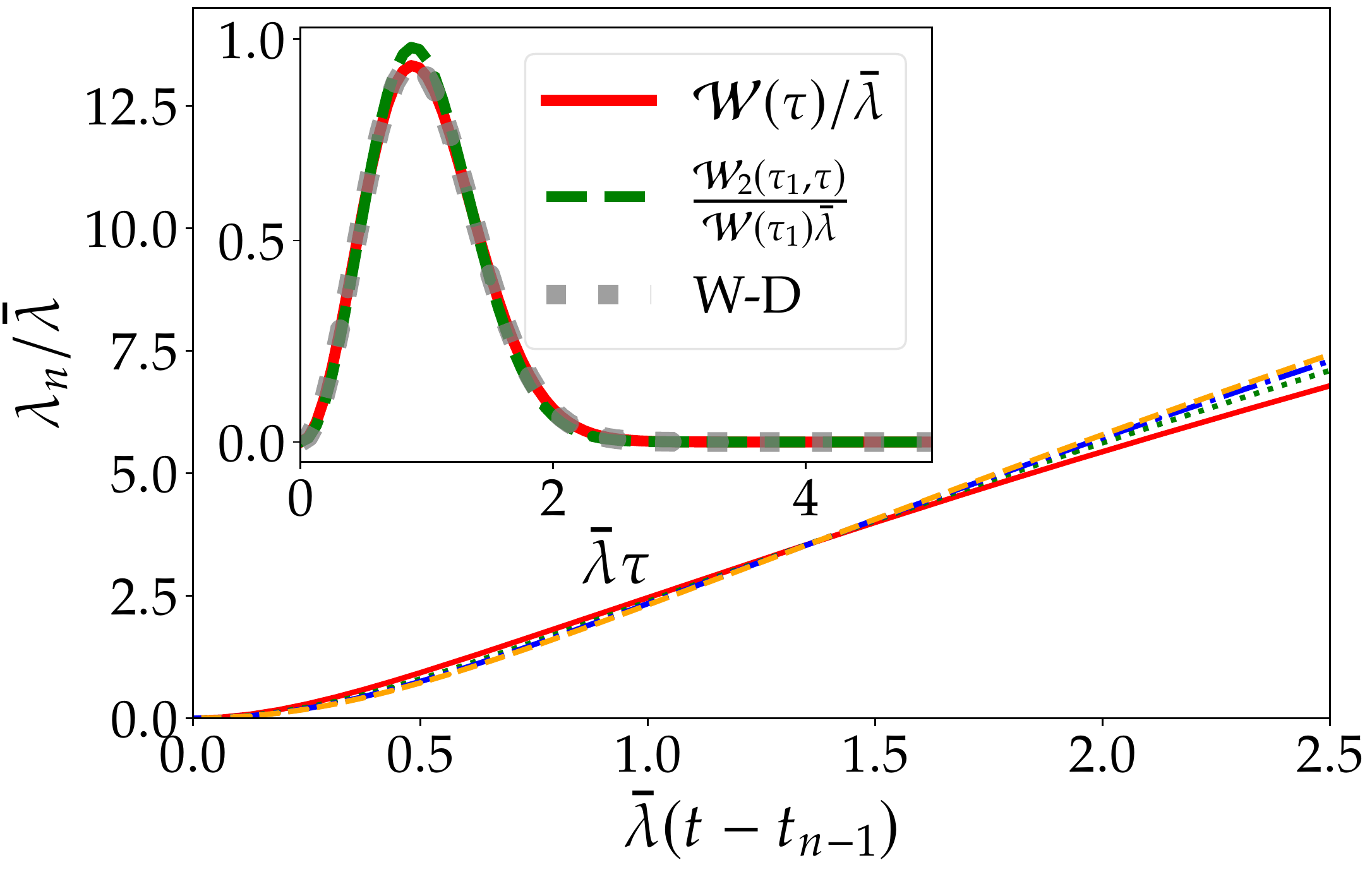}
  \caption{The same as Fig.~\ref{fig5}, but with $D=1.0$. The grey dotted curve
    in the inset corresponds to the Wigner-Dyson distribution.}
  \label{fig11}
\end{figure}

The correlation is most pronounced as the QPC is fully opened. This is
demonstrated in Fig.~\ref{fig10} and~\ref{fig11}, corresponding to $D=1.0$. In
this case, all the conditional emission rates do not saturate at all. They
increase almost linearly as functions of $t$, which is plotted in the main panel
of Fig.~\ref{fig11}. Note that in this case, the corresponding WTD fully agrees
with the Wigner-Dyson distribution, which can be seen by comparing the red solid
curve to the grey dotted curve in the inset of Fig.~\ref{fig11}.

\section{Conclusion}
\label{sec5}

We have shown that the electron emission rate through a quantum point contact
can be described by using the conditional intensity function
$\lambda(t|\mathcal{H}_t)$. For non-interacting systems, the conditional
intensity function $\lambda(t|\mathcal{H}_t)$ can be obtained from the
first-order correlation function. It provides an intuitive way to understand the
temporal behavior of the emission process. As the QPC is close to pinch-off, the
conditional intensity function $\lambda(t|\mathcal{H}_t)$ can be
well-approximated by a constant emission rate $\bar{\lambda}$ in most
regions. This indicates that the emission process can be treated as a simple
Poisson process on long time scales. The correlations between electron emissions
manifests themselves as sharp dips in the conditional intensity function
$\lambda(t|\mathcal{H}_t)$, which can only play a role on short time scales. As
the QPC is opened up, the correlations become more and more important. For QPC
with modest transparency, the dips due to the correlation evolve into wide
valleys, which become non-negligible even on long time scales. In this case, the
emission process can be treated approximately as a renewal process, whose
statistics behaviors can be solely decided via the WTD $\mathcal{W}(\tau)$. For
QPC with high transparency, the correlations are so strong that the conditional
intensity function $\lambda(t|\mathcal{H}_t)$ exhibits a saw-tooth behavior. In
this case, the emission process can only be described within the non-renewal
theory. These results indicates that the conditional intensity function
$\lambda(t|\mathcal{H}_t)$ provides a unified description of the emission
process, which can be used to model both the renewal and non-renewal behaviors.

\bibliographystyle{apsrev4-2}
%

\end{document}